# Viscous motion of spherical nanoparticles that scatter laser radiation in the Rayleigh regime

M.Ya. Amusia[1,2] and A.S. Baltenkov[3]

[1] Racah Institute of Physics, the Hebrew University, Jerusalem, 91904 Israel
[2] Ioffe Physical-Technical Institute, St. Petersburg, 194021 Russia
[3] Arifov Institute of Ion-Plasma and Laser Technologies,
Tashkent, 100125 Uzbekistan

**Abstract.** The mechanism of transverse radiation viscosity for nanospheres moving in laser field is analyzed. It is demonstrated that in the process of light scattering by these particles besides the force $\mathbf{F}_s$ accelerating them in the direction of radiation propagation and the gradient force $\mathbf{F}_g$ that is due to the spatial inhomogeneity of the light field, there are forces $\mathbf{F}_{visc}$ that slow down the movement of particles in the transverse directions. These light viscosity forces are due to the Doppler shift in frequency of scattered radiation. The general expressions for these forces acting on particles that scatter radiation in the Rayleigh regime are derived and applied to estimate their effect on levitated nanospheres and also on slow electrons moving in the laser and magnetic fields. The possibile experiments for observation the effects of light viscosity is discussed.

**1.** Optical levitation of single nano- or microscopic particles in a laser beam is a subject of research in a large number of articles (see for example [1-7] and references therein). The levitation isolates the particles from the environment and provides a platform for precision measurements of weak forces like quantum friction force [8] or the Casimir force [9], if such a particle can be set into rotation at sufficiently high angular velocities and positioned next to a surface [10]. Such a rotational micromanipulation has been achieved with nonspherical or birefringent particles using linearly polarized light [11-13]. In particularly, the optical levitation of silica nanodumbbells in high vacuum was investigated in experiments [14], where by a circularly polarized laser field, the nanodumbbells were accelerated up to the frequency beyond 1 GHz and the linear speed of the nanospheres forming the dumbbell reached $v \approx 10^4$ cm/s. At such speeds of nanosphere motion perpendicular to the light flux (this is exactly what happens in experiments [14]), the light friction forces caused by the Doppler effect can become noticeable.

The origin of these viscosity forces is easy to understand by considering the following *gedanken* experiment. Imagine an ideal pool table, on the surface of which, without friction and energy losses, a billiard ball forever moves. Suppose that the surface of the ball reflects isotropic photons with a frequency $\omega$, a flux of which fall perpendicularly upon the table. Since the ball moves across the light flux, the frequency of the radiation scattered by it is equal to $\omega$. The frequency of the radiation $\omega'$ scattered by the ball moving with the speed **v** relative to the pool table is defined as follows [15]

$$\omega' \approx \omega\left[1+\frac{(\mathbf{v}\cdot\mathbf{n}')}{c}\right]. \qquad (1)$$

Here **n'** is the vector of the scattered photon emission. According to (1), the frequency of the photons emitted forward (in the direction of vector **v**) $\omega' \approx \omega(1+v/c)$ is greater than that of the photons emitted in the opposite direction $\omega' \approx \omega(1-v/c)$. The difference between the impulses of the scattered radiation along the vector **v** and against it is about $\Delta\hbar\omega'/c \approx 2\hbar\omega v/c^2$. This difference is extracted from the momentum of the ball, which slows down its movement across the photon flux.

**2**. One can observe a similar picture in experiments with levitating nanospheres that scatter laser radiation in the Rayleigh mode. It is well known that there are two qualitatively different optical forces that act on a particle in an electromagnetic field [16]. The first of them is called the scattering force $\mathbf{F}_s$. It is proportional to the scattering cross section of the particle and the light beam. For paraxial beams, it is directed along the Poynting vector $\mathbf{S}$ and accelerates a particle in the direction of the light wave propagation. The second one is called gradient force $\mathbf{F}_g$. This force arises in the presence of a spatial gradient of the light beam intensity. It is equal to the Lorentz force acting on the induced dipole moment of the particle, and coincides with the direction of the intensity gradient. In [16], formulas were obtained for the $\mathbf{F}_s$ and $\mathbf{F}_g$ forces acting on a dielectric sphere rested in a laser beam with the Gaussian profile of light intensity. The levitated sphere moving across this beam finds itself under the action of additional *radiative viscous forces* $\mathbf{F}_{\text{visc}}$ that slow down its movement. As far as we know, their influence on the behavior of levitating nanoparticles has never been considered[1]. Below we derive formulas for these forces and discuss the possibilities of their experimental observation.

The differential cross section for Rayleigh scattering of unpolarized radiation by a small sphere into a solid angle $d\Omega'$ around the unit vector $\mathbf{n'}$ reads [19]

$$\frac{d\sigma}{d\Omega'} = \frac{1}{2}\left(\frac{\omega}{c}\right)^4 |\alpha(\omega)|^2 [1+(\mathbf{n}\cdot\mathbf{n'})^2] , \qquad (2)$$

where $c$ is the speed of light, $\mathbf{n}$ is the unit vector of incident wave, $\omega$ is its frequency; $\alpha(\omega)$ is the dynamical polarizability of the sphere. The radius of the sphere is assumed to be small compared with the radiation wavelength $\lambda=2\pi c/\omega$. The total cross section of elastic radiation scattering is given by

$$\sigma(\omega) = \frac{8\pi}{3}\left(\frac{\omega}{c}\right)^4 |\alpha(\omega)|^2. \qquad (3)$$

The scattering force of the light pressure on the sphere is well-known [15]

$$\mathbf{F}_s = \sigma(\omega) W \mathbf{n}, \qquad (4)$$

where $W = \langle E^2/4\pi \rangle$ is the time-averaged energy density of electromagnetic radiation, $\mathbf{E}$ is the electric field of the light wave. Suppose that the energy flux density in the upward direction (along the Z axis) is such that the scattering force of the light pressure $|\mathbf{F}_s|$ is equal to the sphere's weight

$$|\mathbf{F}_s| = \sigma(\omega) W = mg , \qquad (5)$$

where $m$ is the mass of the sphere, $g$ is the acceleration of gravity.

---

[1] Note that at a qualitative level, the effect of light viscosity for atoms in a laser beam was first considered in our note [17]. The effect of light viscosity on fast electrons was also discussed in [18].

Let us consider the 2D-motion of the sphere in the XY plane, Z is the direction of incoming photon beam. As in the case of a billiard ball (1), the frequency of scattered photons $\omega$ is transformed as a result of elastic light scattering into the frequency $\omega'$. In terms of transmission of the scattered radiation impulse, the scattering process can be considered as absorption of photons moving along the vector $\mathbf{n}$ followed by emission of new photons in the direction of the scattered radiation $\mathbf{n'}$. The resultant momentum transferred is the vector of the difference between the momenta of the emitted and incoming photons, respectively. The projection of the photon momentum scattered by the sphere $\hbar\omega'/c$ onto its motion vector $\mathbf{v}$ is equal to $\hbar\omega'(\mathbf{n'}\cdot\mathbf{n''})/c$, where the unit vector is $\mathbf{n''} = \mathbf{v}/v$. Averaging this projection of the photon momentum on all directions of its scattering, we obtain, with the cross section (2), the following expression for the viscous force responsible for slowing the translational motion of the sphere perpendicular to the photon flux (in the XY plane)

$$\mathbf{F}_{visc} = -W\mathbf{n''}\int\frac{d\sigma}{d\Omega'}\left[1+\frac{v}{c}(\mathbf{n'}\cdot\mathbf{n''})\right](\mathbf{n'}\cdot\mathbf{n''})d\Omega' = -\frac{3}{10}\frac{\mathbf{v}}{c}W\sigma(\omega) = -\frac{3}{10}\frac{\mathbf{v}}{c}|\mathbf{F}_s|. \quad (6)$$

Replacing in (6) the cross section for unpolarized radiation (2) by the cross section for the polarized light [19]

$$\frac{d\sigma}{d\Omega'} = \left(\frac{\omega}{c}\right)^4 |\alpha(\omega)|^2 [1-(\mathbf{e}\cdot\mathbf{n'})^2], \quad (7)$$

we obtain the following expression for the viscous force acting on nanosphere moving in the polarized radiation field

$$\mathbf{F}_{visc} = -\frac{1}{5}\frac{\mathbf{v}}{c}[2-(\mathbf{e}\cdot\mathbf{n''})^2]W\sigma(\omega) = -\frac{1}{5}\frac{\mathbf{v}}{c}[2-(\mathbf{e}\cdot\mathbf{n''})^2]|\mathbf{F}_s|. \quad (8)$$

The value of the force (8) depends on the relative orientation of the polarization vector $\mathbf{e}$ and the velocity vector $\mathbf{v}$: it reaches the maximal value when $\mathbf{e}$ is perpendicular to $\mathbf{v}$ and the minimal value when $\mathbf{e}\|\mathbf{v}$. The magnitude of forces (6) and (8) make up certain fractions of the force (4), and these fractions are completely determined by the Rayleigh differential scattering cross sections (2) and (7). The motion of the nano-spheres under the action of these forces resembles their motion in a viscous medium under the action of the Stokes' force.

Let us estimate the value of the viscosity force (6) for a single levitated nano-sphere. The scattering force of the light pressure, produced by the laser in [14] with the power $P = 0.5$ W and wavelength $\lambda = 1550$ nm (tightly focused on the sphere) is $F_s = P/c = 1.7\cdot10^{-9}$ N. The linear speed of a nanodumbbell sphere in the experiment [14] $v \approx 85$ nm·1 GHz $= 8.5\cdot10^3$ cm/s. Using this speed value, we obtain $F_{visc} = 1.4\cdot10^{-16}$ N for the force (6). This force exceeds the zeptonewton forces investigated in paper [20] by five orders of magnitude.

**3**. As a possible experiment for observing the forces of light friction, we consider the rotation of a nanodumbbell that scatters laser radiation in the Rayleigh mode. Consider a levitated nanodumbbell not in the form of two sticking together balls, as in [14], but in the form of the Cavendish torsion balance with two spheres connected by a handle with a length $l$ (see dimer of nanospheres in [12]). We assume also that the handle of our dumbbell is a thin enough rod, the viscous friction of which in the light flux can be neglected. Let the length of the dumbbell handle be $l \gg \lambda$, so that the scattering of radiation by dumbbell balls can be considered in the Rayleigh (Rel) regime. Suppose that the laser beam is initially linearly polarized and the dumbbell is captured by optical tweezer, so that the middle of the rod is in the center of the laser beam. Therefore, the sum of the gradient forces $\mathbf{F}_g$ acting on the dumbbell balls is equal to zero. As in [14], the dumbbell is accelerated by circularly polarized radiation up to the angular speed $\dot{\varphi}_0$ [2]. After that, the laser radiation is instantly converted to the non-polarized one. As a result, rotation of the dumbbell in the XY plane slows down under the action of the viscous force (6). The following simple equation describes this process

$$m\frac{l^2}{2}\ddot{\varphi} = -\frac{3}{10}\frac{2mg}{c}\frac{l^2}{2}\dot{\varphi}. \tag{9}$$

Here $ml^2/2$ is the dumbbell moment of inertia, $2mg$ is the scattering force (5), which provides dumbbell's levitation in the light beam; $\ddot{\varphi}$ is the angular acceleration of the dumbbell axis. The solution to this equation is obvious: $\dot{\varphi}(t) = \dot{\varphi}_0 \exp(-t/\tau_R)$. Here

$$\tau_{\mathrm{Re}l} = 5c/3g \approx 5.1 \times 10^7 \text{ sec} \tag{10}$$

is the relaxation time, i.e. the period of time required for angular (or linear) speed reduction of the levitated nanodumbbell by a factor of about of 3. It is remarkable that this characteristic time for the Rayleigh mode does not depend on the size and mass of the levitating dumbbell, but is determined only by the radiation scattering cross section (6).

The Rayleigh regime of radiation scattering suggests that the EM-wavelength is much larger than the radius of nanosphere $a$ ($\lambda >> a$). Consider, as in the case of above-investigated billiard ball) another limiting case: when the wavelength $\lambda$ is small compared to the radius of the spheres $a$ forming the nanodumbbell. We will consider these spheres to be perfectly conducting and, thus, perfectly reflecting. The scattering of radiation by them can be considered in the approximation of geometric (Geom) optics, in the framework of which the differential cross section for scattering of photons by a sphere is [21]

$$\frac{d\sigma}{d\Omega'} = \frac{a^2}{4}. \tag{11}$$

Substituting this cross section into formula (6), we obtain the following expression for the viscosity force of a nanosphere in the light flux $W$

$$\mathbf{F}_{visc} = -W\mathbf{n}''\frac{a^2}{4}\int\left[1+\frac{v}{c}(\mathbf{n}'\cdot\mathbf{n}'')\right](\mathbf{n}'\cdot\mathbf{n}'')d\Omega' = -\frac{1}{3}\frac{\mathbf{v}}{c}\pi a^2 W = -\frac{1}{3}\frac{\mathbf{v}}{c}|\mathbf{F}_s|. \tag{12}$$

For the time of slowing down the speed of the levitated nano-dumbbell rotation, we obtain instead of (10) the following result:

$$\tau_{\mathrm{Geom}} = 3c/2g \approx 4.6 \times 10^7 \text{ sec}. \tag{13}$$

The given above values of time $\tau_{\mathrm{Rel}}$ and $\tau_{\mathrm{Geom}}$ can be done smaller if we increase the radiation density $W$. We add to the first laser beam a second one, directed vertically down along the Z axis. Let us suppose that the powers of the laser beam-1 (with $W_1$) and the laser

---

[2] A dot symbol at the angular speed denotes differentiation with respect to time.

beam-2 (with $W_2$) are such that the nanogumbbell, as before, levitates in the optical cavity created by two counter-propageted laser beams (see Fig. 1(b) in [1]). In this experimental device, the scattering pressure forces, $\mathbf{F}_s^{(1)}$ and $\mathbf{F}_s^{(2)}$, are subtracted, while the viscosity forces $\mathbf{F}_{visc}$ are summated. It follows from the fact that the viscosity forces are determined only by the density of the electromagnetic radiation $W$, i.e. by the number of photons per unit volume, regardless of the direction of photon motion. In this case in formulas (6) and (12) we have to replace $W$ by epy sum of densities $W_1+W_2$. The increase in photon density is limited, of course, by the laser induced heating and evaporation of the nanospheres.

**4.** Consider a nanoobject that scatters laser radiation in the Rayleigh regime, for which the problem of heating and evaporation does not exist. Suppose the optical cavity formed by the two laser beams, considered above, is placed in a strong magnetic field $\mathbf{B}$ along the Z axis. A slow electron with speed $v$ is injected into this cavity perpendicular to the Z axis and begins to move along a circular trajectory with radius $R = vm/eB$ in the XY plane. Here $e$ and $m$ are the electron charge and mass, respectively. Let us also suppose that the radius of the optical cavity $w_0$ is slightly greater than the cyclotron radius, $w_0 \geq R$. Under this condition, the circular orbit of the electron remains inside the light spot. The equations of motion of an electron in light and magnetic fields are determined by the following formulas

$$m\dot{v}_x = -ev_y B - \beta v_x,$$
$$m\dot{v}_y = ev_x B - \beta v_y. \quad (14)$$

The parameter $\beta$ in (14) is described by the following equation

$$\beta = \frac{3}{10}\frac{W\sigma}{c}. \quad (15)$$

Here $\sigma$ is the Thomson cross section of elastic light scattering [15]

$$\sigma = \frac{8\pi}{3}\left(\frac{e^2}{mc^2}\right)^2. \quad (16)$$

Multiplying Eqs.(14) by $v_x$ and $v_y$, respectively, and subtracting them, we obtain equation $\dot{v} = -(\beta/m)v$, where $v^2 = v_x^2 + v_y^2$. The solution of this equation is the following

$$v(t) = v_0 e^{-\frac{\beta}{m}t}. \quad (17)$$

The time-parameter in the exponent (17) is determined by the following expression

$$\tau = \frac{m}{\beta} = \frac{5}{3}\frac{\pi w_0^2}{\sigma}\frac{mc^2}{P}. \quad (18)$$

We assume in Eq. (18) that the powers $P$ of the beam-1 and the beam-2 are the same. The magnitudes of the longitudinal forces of light pressure, directed toward each other, are, therefore, equal, while the density of electromagnetic radiation $W$ is increased by a factor of two. For a slow electron with energy $\varepsilon \approx 30$ eV in the magnetic field $B = 1$ T, the radius of the cyclotron orbit is $R \approx 1.85 \cdot 10^{-5}$ m. For an estimate, let us substitute $w_0 = 2 \cdot 10^{-5}$ m into Eq. (18). At the beam power $P=0.5$ kW, the experiment time, required for the initial electron velocity $v_0 \approx 3.25 \cdot 10^6$ m/s to decrease by a factor of three is $\tau \approx 0.5 \cdot 10^4$ s. The spiral path of the electron in the superposition of the magnetic and viscous light fields is presented in the figure 1. The magnetic fields $B$ and the electron energy $\varepsilon$ used in these estimates are experimentally quite accessible.

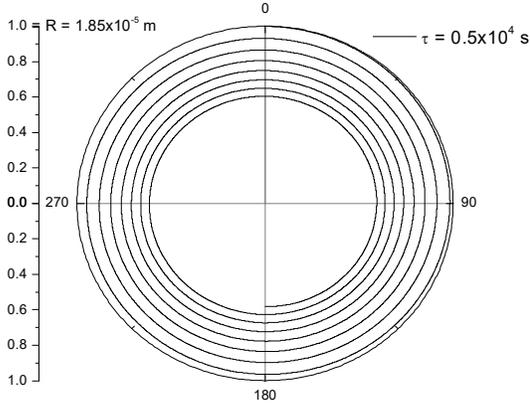

Fig.1. Electron trajectory in magnetic and laser fields calculated with Eq. (14) for initial radius $R \approx 1.85 \cdot 10^{-5}$ m and time $\tau \approx 0.5 \cdot 10^4$ s.

**4.** We presented here formulas that permit to calculate the radiative viscosity and investigate its specific features. The force, quite expectedly, proved to be weak, but in principle influential and distinguishable from other forces by which a light beam acts upon nanoobjects and electrons. Recently developed optomechanic experimental technique makes them observable. We suggested some effects that could unveil this force. We hope that the observation of these new physical manifestations of light viscosity as well as confirmation of their predicted values would be of great interest.

**Acknowledgments**
ASB is grateful for the support to the Uzbek Foundation Award OT-Ф2-46 and Dr. A. V. Zinoviev for useful discussions.

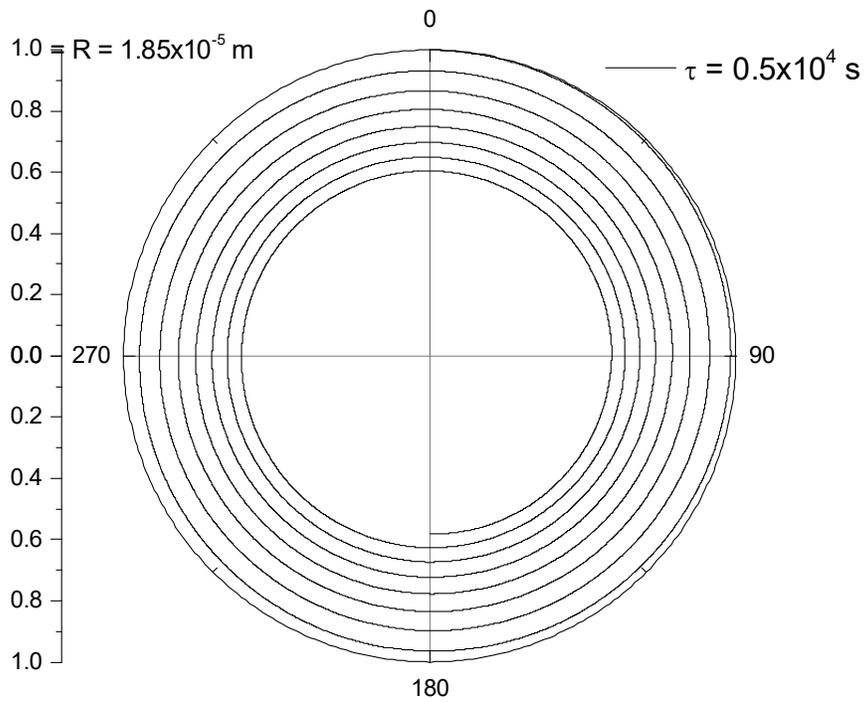

Fig.1. Electron trajectory in magnetic and laser fields calculated with Eqs.(14) for initial radius $R \approx 1.85 \cdot 10^{-5}$ m and time $\tau \approx 0.5 \cdot 10^{4}$ s.